\documentclass[preprint,12pt]{elsarticle}
\usepackage{amssymb}
\usepackage{graphicx}
\usepackage{amsmath,amsfonts}
\usepackage{xtab}

\date{ }

\begin{document}

\begin{frontmatter}

\title{Simulated Annealing for Optimal Ship Routing}

\author[UoP]{O. T. Kosmas}
\ead{odykosm@uop.gr}

\author[UoP]{D. S. Vlachos}
\ead{dvlachos@uop.gr}

\address[UoP]{Laboratory of Computer Sciences,\\
Department of Computer Science and Technology,\\
Faculty of Sciences and Technology, University of Peloponnese\\
GR-22 100 Tripolis, Terma Karaiskaki, GREECE}

\begin{abstract}
A simulated annealing based algorithm is presented
for the determination of optimal ship routes through the
minimization of a cost function. This cost function is a weighted
sum of the time of voyage and the voyage comfort (safety is taken
into account too). The latter is dependent on both the wind speed
and direction and the wave height and direction. The algorithm first
discretizes an initial route and optimizes it by considering small
deviations which are accepted by utilizing the simulated annealing
technique. Using calculus of variations we prove a key theorem which
dramatically accelerates the convergence of the algorithm. Finally
both simulated and real experiments are presented.
\end{abstract}


\end{frontmatter}

\section{Introduction}
\label{}
Optimization of ship routing is closely related to both ship
characteristics and environmental factors. Ship and cargo
characteristics have a significant influence on the application of
ship routing. Ship size, speed capability and type of cargo are
important considerations in the route selection process prior to
sailing and the surveillance procedure while underway. A ship's
characteristics identify its vulnerability to adverse conditions and
its ability to avoid them \cite{vla2}.

On the other hand, environmental factors of importance to ship
routing are those elements of the atmosphere and ocean that may
produce a change in the status of a ship transit. In ship routing
consideration is given to wind, waves, fog and ocean currents. While
all of the environmental factors are important for route selection
and surveillance, optimum routing is normally considered attained if
the effects of wind and waves can be optimized. The effect of wind
speed on ship performance is difficult to determine. In light winds
(less than 20 knots) ships lose speed in headwinds and gain speed
slightly in following winds. For higher speeds, ship speed is
reduced in both head and following winds. Wave height is the major
factor affecting ship performance. Wave action is responsible for
ship motions, which reduce propeller thrust and cause increased drag
from steering corrections. The relationship of ship speed to wave
direction and height is similar to that of wind. Head seas reduce
ship speed, while following seas increase ship speed slightly to a
certain point, beyond which they retard it. In heavy seas, the exact
performance may be difficult to predict because of the adjustments
of the course and speed for ship handling and comfort. Although the
effect of sea and swell is much greater than wind, it is difficult
to separate the two in ship routing. Fog, while not directly
affecting ship performance, should be avoided as much as feasible,
in order to maintain normal speed in safe conditions. Although the
route may be longer by avoiding fog, transit time may be less due to
not having to reduce speed in reduced visibility. In addition, crew
fatigue due to increased watch keeping vigilance can be reduced.
Ocean currents do not present a significant routing problem, but
they can be a determining factor in route selection and diversion.
The important consideration to be evaluated are the difference in
distance between a great circle route and a route selected for
optimum current, with the expected increase of speed of advance from
the following current. More details about the effect of
environmental factors can be found in \cite{vla1}.

The role of routing is to allocate the available resources to best
fulfill certain requirements. In this paper we are interested in the
problem of optimal ship routing taking into account only the wave
height and direction by using the simulated annealing algorithm
\cite{kir1,kir2}. The simulated annealing method is an extension of
a Monte Carlo method developed by Metropolis et al \cite{met}, to
determine the equilibrium states of a collection of atoms at any
given temperature T. Since the method was first proposed in
\cite{kir1,kir2}, much research has been conducted on its use and
properties \cite{bro,cha,cor,dar,ing1,ing2,nah,rav}. The method
itself is a technique which has attracted significant attention as
suitable for optimization problems of large scale. It can give
solution when a desired global extremum is hidden among many,
poorer, local extrema. Even though other practical methods have also
been found, surprisingly, the implementation of the algorithm is
relatively simpler. At the heart of the method is an analogy with
thermodynamics, specifically with the way liquids freeze and
crystallize, or metals cool and anneal, when they are cooled slowly
and thermal mobility is lost. For slowly cooled systems, nature is
able to find the minimum energy state.

The paper is organized as follows: in section 2 the formulation of
the problem is given. In section 3 we present the method of discrete
variational principles and we prove a key result which allows us for
an efficient way of searching for the optimal solution. In section
4 the steps of the algorithm are explained and analyzed. Finally
in section 5 both simulated and real experiments are performed and
the results are presented and discussed.

\section{Formulation of the problem}
\begin{figure}
\begin{center}
\includegraphics[width=\textwidth]{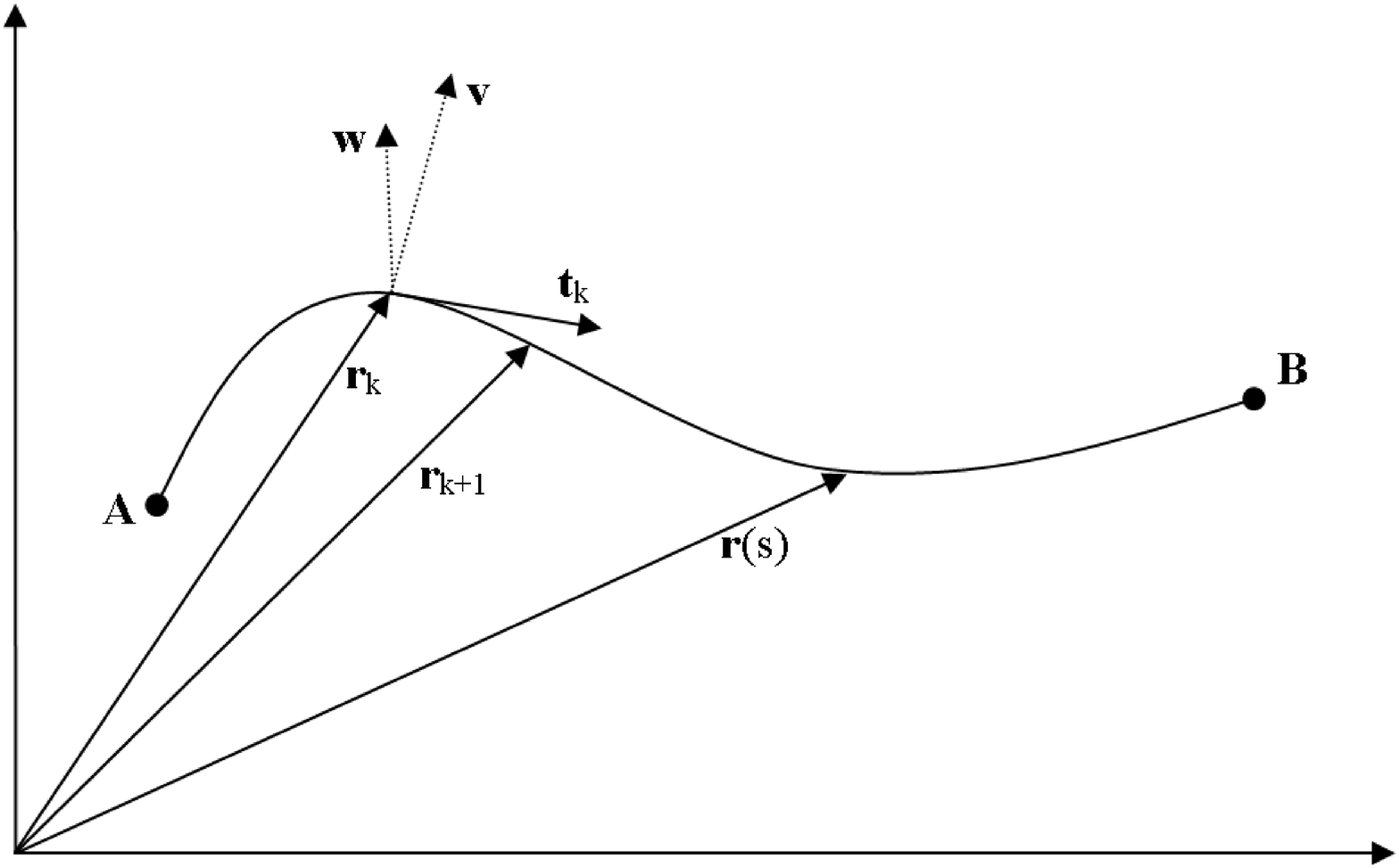}
\end{center}
\caption{A route from point A to point B. $r(s)$ is the
parametrization of the route, and $w,v$ are the wave and wind
vectors respectively.}\label{fig-form}
\end{figure}

Let us assume that an initial route of a ship is represented by a
smooth curve $\vec{r}(s)$  (fig. \ref{fig-form}), where the parameter
$s$ is the arc length measured from some fixed point A (initial
point of the ship route). Then, the tangent vector of the curve of
the ship's route in the point of question (see e.g. \cite{las}) is
defined as
\begin{equation}
\vec{t}=\frac{\dot{\vec{r}}}{|\dot{\vec{r}}|}=\frac{d\vec{r}}{ds}
\end{equation}
where  $\dot{\vec{r}}$ is the ship's velocity.

We also assume that the moving ship is subject to the influence of
the wave height and direction represented by the vector $\vec{w}$
and the wind speed and direction represented by the vector
$\vec{v}$.

Under the above assumptions, we define a route cost (a scalar
quantity) assigned at every possible route between two points. The
route cost includes a weighted combination of the voyage time and
the safety (or comfort) of the voyage. The total cost $S$ is given
by
\begin{equation}
S=aT+(1-\alpha )C
\end{equation}
where $T$ is the total voyage time and $C$ is a scalar
characterizing the safety (comfort) of the voyage. The weight
$\alpha$ can be tuned by the user depending on his demands. Note
here that when $\alpha = 1$, then the only optimization parameter is
the voyage time while when $\alpha = 0$, the only optimization
parameter is the crew comfort. The scalar $C$ is calculated as a
line integral over the route by the following way (up to the linear
approximation):
\begin{equation}
C=\int_{A}^{B}{(\vec{v}^{T}Z_{v}+\vec{w}^{T}Z_{w})\vec{t}ds}
\end{equation}
where $\vec{v}$ is the wind vector, $\vec{w}$ is the wave height
vector and $Z_{v}$ and $Z_{w}$ are tensors which characterize the
ship response to wind and wave, respectively. The calculation of the
total voyage time $T$ is a bit more complicated, since both wind
and waves can alter the speed of the ship. In general we can write
that
\begin{equation}
|\dot{\vec{r(r)}}|=u+F(\vec{v},\vec{w},\vec{t})
\end{equation}
where $u$ is the speed of the ship in zero wind and $F$ is a
function that depends on ship characteristics, wind, wave and
direction of the ship movement. For simplicity in the present work,
we assume that $F = 0$.

\section{Discrete variational principles for mechanical problems}
Recently, progress has been made in the development of variational
discretization in mechanical problems, both in the fundamental
theory and in the applications to challenging problems. Among them,
a method of variational integrators based on the discretization of
Hamilton's principle has been developed which underlies essentially
all of mechanics, from particle mechanics to continuum mechanics.
Also the discretization of Lagrange-D'Alembert principle in cases
where there is dissipation or external forces present has been used.

Preserving the basic variational structure in the algorithm retains
the structure of mechanics (such as conservation laws) at the
algorithmic level. This avoids many of the problems with existing
integrators, such as spurious dissipation, which may take very
expensive runs to eliminate by standard techniques.

With an appropriate development of the connection between mechanics
and geometry in the discrete setting, one will be able to use for
geometry the technology described in Ref. \cite{des}. The theory of
variational integrators is originated from the Hamilton-Jacobi
theory and parts of the basic theory have been constructed in Ref.
\cite{mos} and its numerical analysis is due to various groups,
\cite{wen1,wen2}.

In the first step of the present work we construct the comfort
function C which must be treated with minimization techniques as a
basic problem of the calculus of variation. This problem resembles
to that appeared in Hamilton's principle where a Lagrangian
$L(q,\dot{q})$ is required.  Namely to make the integral
\begin{equation}
\delta C=\delta \int_{A}^{B}f(q,\dot{q})dt=0
\end{equation}
stationary for all variations $\delta s$ of $s$ that vanish at the
fixed end points, which leads to the Euler-Lagrange equations. The
curve obtained by projecting this solution q onto the shape, also
solves a variational principle on the shape space. This theory has a
PDE counterpart in which one makes a space-time integral stationary
which is appropriate for elasticity and fluids for example.

The benefit when the variational integrators derive from the discrete variational mechanics
can be shown by reviewing the derivation of
the Euler-Lagrange equations and mimic this progress on a discrete
level.

\subsection{Discretization of the initial ship's route}
We consider now the discrete case, where the route is approximated
by a polygonal line with edges $\vec{r}_{k},k=0,1,....M$ . The
distance of each line segment is
\begin{equation}
\delta _{k}=|\vec{r}_{k}-\vec{r}_{k+1}|,k=0,1,....M-1
\end{equation}
and the tangent vector at each edge is
\begin{equation}
\vec{t}_{k}=\frac{\vec{r}_{k}-\vec{r}_{k+1}}{\delta
_{k}},k=0,1,....M-1
\end{equation}
The cost function can be calculated then as:
\begin{eqnarray}
\nonumber T & =& \sum_{k=0}^{M-1}\frac{\delta _{k}}{u} \\
\nonumber C & =&
\sum_{k=0}^{M-1}\vec{z}_{k}^{T}\vec{t}_{k}\delta_{k} \\
 S & =& aT+(1-a)C
\end{eqnarray}
where
\begin{equation}
\vec{z}_{k}^{T}=(\vec{v}^{T}Z_{v}+\vec{w}^{T}Z_{w})
\end{equation}
calculated at point $\vec{r}_{k}$ and at time
\begin{equation}
T_{k}=\sum_{j=0}^{k-1}\frac{\delta _{k}}{u}
\end{equation}
Consider now a small change in $\vec{r}_{k}$ by $\epsilon
\vec{x}_{\phi}$, where $\epsilon$ is a small positive number and
$\vec{x}_{\phi}$ is the unit vector with angle $\phi$ with the
horizontal axis. The new route will have cost $S'$ with
\begin{eqnarray}
\nonumber \Delta S & =& S'-S \\
\nonumber  & =&  a\epsilon
(\vec{z}_{k}^{T}-\vec{z}_{k-1}^{T})\vec{x}_{\phi}+(1-a)\epsilon
\frac{(\vec{t}_{k}^{T}-\vec{t}_{k-1}^{T})\vec{x}_{\phi}}{{u}} \\
& =& \epsilon \Bigl(a(\vec{z}_{k}^{T}-\vec{z}_{k-1}^{T})+
(1-a)\frac{\vec{t}_{k}^{T}-\vec{t}_{k-1}^{T}}{u}\Bigr)\vec{x}_{\phi}
\end{eqnarray}
where we have assumed that
\begin{equation}
|\vec{r}+\vec{\lambda}|\approx|\vec{r}|
+\frac{\vec{r}\vec{\lambda}}{|\vec{r}|}\;,\;|\vec{\lambda}|\ll|\vec{r}|
\end{equation}
\subsection{Searching using simulated annealing}
Consider now the initial route in Fig. 1. The route is broken into
several line segments, the number of which depends on the solution
of the wind and wave forecasts. In this way a route is represented
as a list of way-points. Initially the total cost of the route is
calculated. Then in every iteration step every way-point of the
route is moved by an elementary length, perpendicular to the line
which connects the departure and the arrival points. This elementary
length is specified by the resolution of the wind and the wave
forecasts. Every movement has a positive or negative contribution to
the total cost. A movement is accepted even if it has a positive
contribution to the total cost with a probability which depends on
the temperature of the system, the so-called Boltzmann probability
distribution,
\begin{equation}
Prob(E) \approx exp(-E/kT)
\end{equation}
expressing the idea that the system in thermal equilibrium at
temperature $T$ has its energy probabilistically distributed among
all different energy states $E$. Even at low temperature, there is a
change, albeit very small, of a system being in a high energy state.
Therefore, there is a corresponding change for the system to get out
of a local energy minimum in favour of finding a better, more global
one. The quantity k (Boltzmann's constant) is a constant of nature
that relates temperature to energy. In other words, the system
sometimes goes uphill as well as downhill, but the lower the
temperature, the less likely is any significant uphill excursion.

\begin{figure}
\begin{center}
\includegraphics[width=\textwidth]{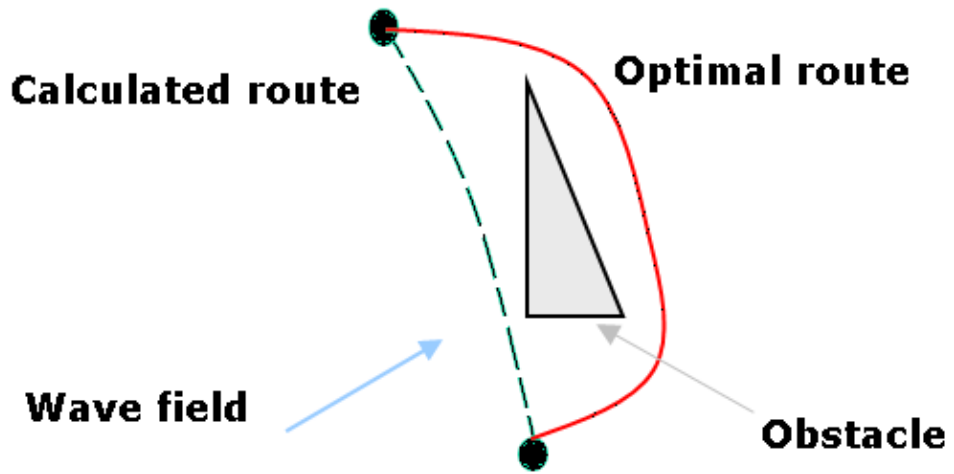}
\end{center}
\caption{The solution can be locked in a local minimum if the
optimal route cannot be generated by continuous transformations of
the initial route.}\label{fig-obs}
\end{figure}

Initially the temperature is high, but as the algorithm proceeds,
the temperature is decreased to zero. At this point, only movements
with negative contributions are accepted. This method is known as
simulated annealing and is used to avoid the local minima. Notice
here that there is no systematic way to decide if the calculated
route is the optimal one. In most cases however, the voyage time is
very critical and thus the optimal route is close to the shortest
initial one. The only case that was observed in our experiments,
where the iterative method was locked in a local minimum, was when
the line between the departure and arrival points was very close to
a small obstacle (fig. \ref{fig-obs}). In this case the method could
not overcome the obstacle and the algorithm was terminated. The reason
is that only continuous transformations of the route are performed
in every step of the algorithm. To overcome this difficulty, we
consider several initial routes, as it will be explained in the next
section.

To accelerate the convergence of the searching mechanism, it would
be useful to know, in which directions, the cost function is more
sensitive to the transformations of the initial route. Fortunately
we can prove that when minimizing the cost function for any segment
of the ship's route, the resulting movement is parallel or
anti-parallel to the direction of the wave propagation. The proof
proceeds as follows (considering only the contribution of the wave
to the cost function). The cost function is given
\begin{equation}
C=\int_{A}^{B}(\vec{w}^{T}Z_w) \vec{t} ds
\end{equation}
and we assume a change to the route close to point $\vec{r}(s_0)$,
which results in a new tangent vector of the form
\begin{equation}
\vec{t}'=\vec{t}+\epsilon \delta(s-s_0) \vec{x}_{\phi}
\end{equation}
where $\delta$ is the Dirac function and $\epsilon,\vec{x}_{\phi}$
are defined in section 3.1. The variation in the cost function will
be
\begin{equation}
\delta C=\int _A^B (\vec{w}^TZ_w)\epsilon
\delta(s-s_0)\vec{x}_{\phi}ds
\end{equation}
or
\begin{equation}
\delta C=\epsilon \vec{z}_{|s=s_0}\vec{x}_{\phi}
\end{equation}
where $\vec{z}_{|s=s_0}$ is the vector $\vec{w}^TZ_w$ calculated at
the point of the route where $s=s_0$. It is obvious now that the
maximum change of the cost will happen if the angle between the
vector $\vec{z}_{|s=s_0}$ and $\vec{x}_{\phi}$ is $0$ or $\pi$. The
acceleration now of the convergence is obtained by selecting
variations of the initial route only in the above directions. In our
experiments, the direction of the vector $\vec{z}_{|s=s_0}$ is
always the same as the direction of the wave propagation, hence the
accepted variations of the initial route are those which are
parallel or antiparallel to the direction of the wave.

\section{Description of the algorithm}
For the initialization of the algorithm, we have to calculate one or
several initial routes. In the case that there are no obstacles
between the starting and ending points, the initial route is simply
the largest cycle joining these two points. On the other hand, if
there is an obstacle between the starting and ending point, then
there are two possible initial routes, each of which bypasses the
obstacle from a different side. It is a good practice to start with
shortest initial routes (one route for every possible bypass), thus
having already optimized the voyage time. In this case it is obvious
that, if point $B$ belongs to the shortest path between points $A$
and $C$, then the shortest path between points $B$ and $C$ is part
of the shortest path between points $A$ and $C$. It is proved in
\cite{vla2} that the shortest path bypassing an obstacle is tangent
to the convex hull of the obstacle. So the problem of calculating
the shortest path, is reduced to the calculation of the shortest
path from the contact point of this tangent to the ending point.
Recursively we get that the number of initial paths is $2^n$, where
$n$ is the number of obstacles between the starting and ending
points. The algorithm must check all the initial routes to decide
for the optimal solution even if the computational time is
increased.

As soon as one initial route is calculated, we have to decide for
the number of way-points that will be used to represent the route.
We can think this number as the degree of a polynomial used to fit a
given smooth curve. Increasing the degree of the polynomial results
in a better fitting. However at the same time there are at least three
reasons to argue that this not exactly the case here. The first one
is that increasing the number of way-points, the computational time
is increased and the actions that have to be taken by the ship's
crew are increased too. The second reason concerns the period
$T_{d}$ of the forecast data. If $u$ is the speed of the ship, then
nothing is known about the drift of the environmental data in the
time interval $T_d$ and thus in the space interval $u\cdot T_d$. The
last reason is the details of the model used to forecast the
environmental data, and more precisely the mesh that is used. If
$L_d$ is the distance of the mesh points, then the ship for the time
interval $L_d/u$ is moved in constant environmental parameters.
Experiments performed with various numbers of way-points showed that
an optimal selection for the distance of the way-points should be
close to $L_d$. It is then a good practice to continuously adjust
the number of way-points. If the distance between two successive
way-points is greater than $D_{M}$, another way-point is added
between them. On the other hand, if two successive way-points are
closer than a special distance $D_{m}$ , the two way-points are
merged into one. In the experiments that will be presented in this
paper
\begin{equation}
\nonumber D_M=2\cdot L_d \;,\; D_m=L_d/2
\end{equation}
The calculation now of the optimal route is performed as follows:
\begin{enumerate}
\item Consider the set $\{\vec{r}_{k},k=0,(1),M\}$ of $M+1$ points describing the
initial route.

\item Choose a random $k$ (between $1$ and $M-1$,
i.e. including all points of the set except for the end points $k=0$ and
$k=M$) and generate a random real $\phi$ such that $0\le \phi \le
2\pi$.

\item Calculate the difference of the cost function
between the initial route and the route resulting by shifting
$\vec{r}_{k}$ by $\epsilon{\vec{x}_{\phi}}$, where $\epsilon$ is a
small real number and $\vec{x}_{\phi}$ is a unit vector in the
chosen random direction.

\item Generate a random number $p$ between $0$ and
$1$ and accept the variation of the route if $p<e^{-\Delta{S}/E}$.
If the variation is not accepted, goto to step 6.

\item Check $\vec{r}_{k}$ with the adjacent points. This means that, if two
points are very close, we merge them or if they are too far, we add a
new point between them.

\item Decrease the temperature $E$ of
the system in analogy with the cooling of the system (simulated
annealing). Check if the algorithm has to terminate, otherwise goto
step 2.
\end{enumerate}

The symbols used above represent the following: $p$ stands for the
probability Boltzmann distribution and $E$ is the temperature of the
algorithm. Other symbols have been explained previously.

\section{Numerical Results}
\subsection{{Constant wave field}}

\begin{figure}
\begin{center}
\includegraphics[width=\textwidth]{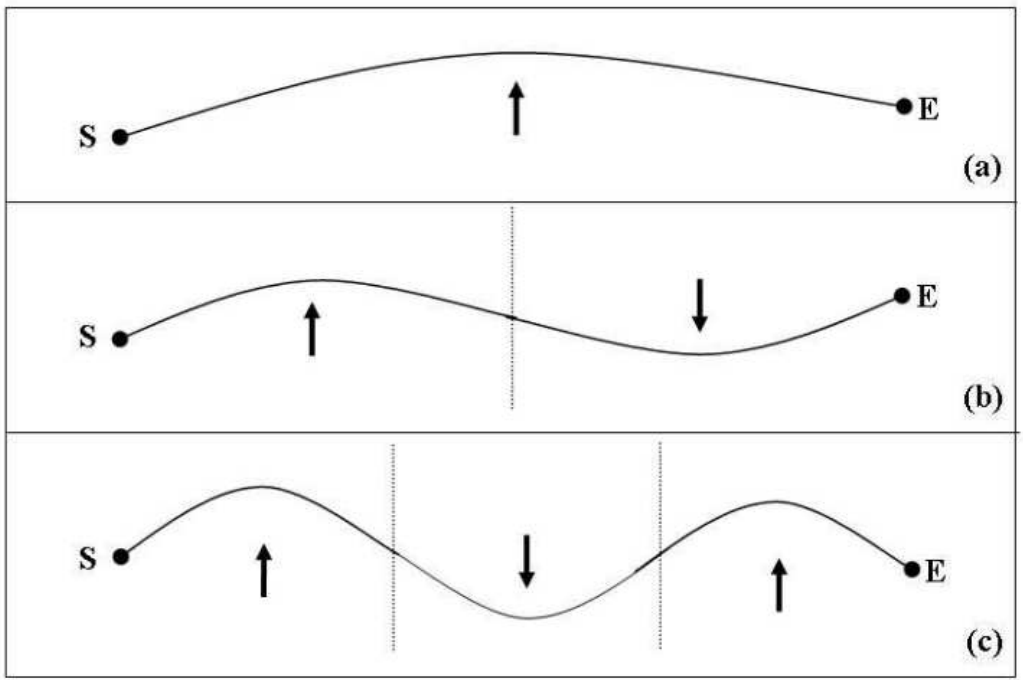}
\end{center}
\caption{Optimal route calculated in (a) a wave field with constant
magnitude and direction and  (b),(c) in a wave field with constant
magnitude but with direction which is inverted once and twice
respectively during the voyage.}\label{fig-exp1}
\end{figure}

In the first experiment (fig. \ref{fig-exp1}), the optimal route for
a uniform (a) or piecewise uniform (b,c) wave field is calculated.
In the first experiment, the direction of the wave was upwards and
does not change through the voyage. Notice that the way-points are
moved upwards in order to maximize the product $|\vec{u}\cdot
\vec{w}|$ , where $\vec{u}$ is the velocity of the ship. This is
very reasonable, since the higher moments around the ship's axis are
developed when the wave is perpendicular to the ship's direction.
The same behavior is observed in fig. \ref{fig-exp1}.b and fig.
\ref{fig-exp1}.c where the magnitude of the wave field is constant
but the direction is inverted once and twice respectively.

\subsection{{The effect of parameter $\alpha$}}
\begin{figure}
\begin{center}
\includegraphics[width=\textwidth]{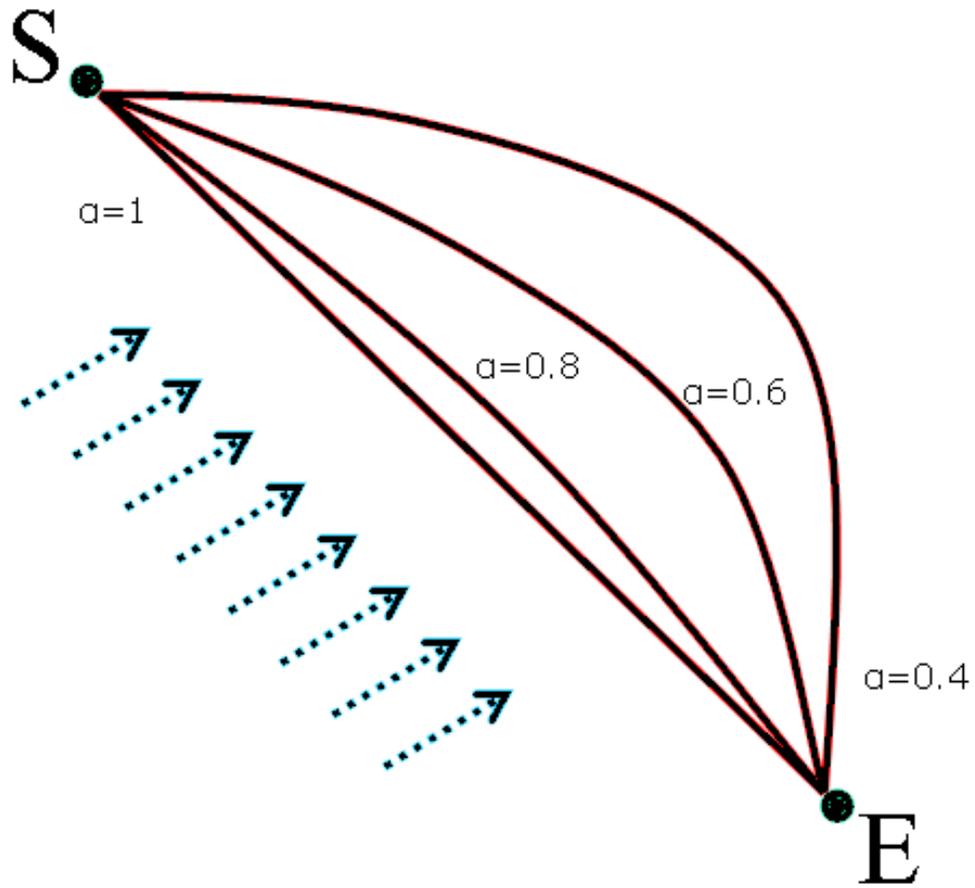}
\end{center}
\caption{The optimal route for various values of $\alpha$. $\alpha=1$ represents the route with minimum route time, while smaller values of $\alpha$ represent a more safe (comfort) voyage.}\label{fig-exp2}
\end{figure}

The effect of the parameter $\alpha$ is studied in the next
experiment (fig. \ref{fig-exp2}). In this experiment, the wave field
is constant with direction perpendicular to the line joining the
starting ($S$) and ending ($E$) point of the voyage. When the
parameter $\alpha$ takes its largest value ($1$), no action is taken
from the algorithm take into account the comfort of the voyage. Since
voyage time is the only parameter to optimize, the calculated value
is simply the largest circle joining the starting and ending points.

By decreasing the value of the parameter $\alpha$, the comfort of
the voyage starts to act cumulatively to the total cost function,
thus making the algorithm to turn the ship parallel or antiparallel
to the direction of the wave field. This tendency is compensated by
the time of the voyage which is increased. By decreasing more the
value of the parameter $\alpha$, the contribution of the voyage time
to the total cost is decreased, leading to longer routes which are
aligned with the direction of the wave field.

\subsection{{Field test}}
\begin{figure}
\begin{center}
\includegraphics[width=\textwidth]{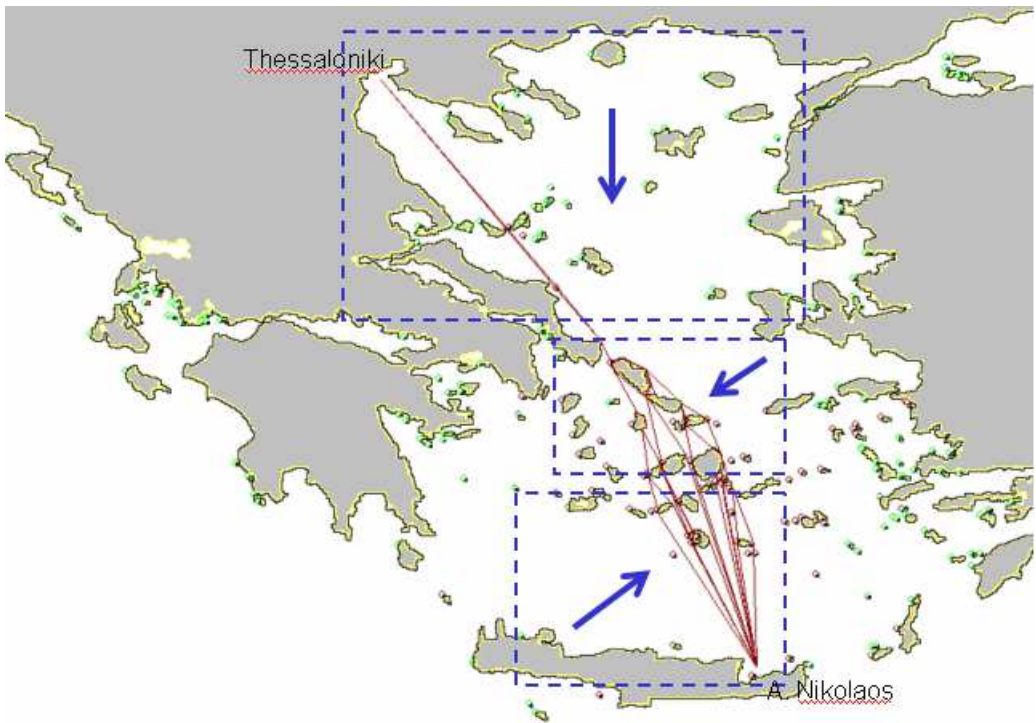}
\end{center}
\caption{Initial routes calculated from the port of Thessaloniki to
the port of A. Nikolaos. The dominant wave fields are shown across
the routes with an arrow showing the direction and magnitude of the
wave field.}\label{fig-init}
\end{figure}

In the next experiment, the optimal route is calculated from the
port of Thessaloniki ($40.5197N,\;22.9709E$) to the port of Ag.
Nikolaos $(35.1508N,$ $25.7227E)$. The forecast data are taken from
the \emph{POSEIDON} system which uses floating buoys for real time
measurements and mathematical models to predict the wave
characteristics for the next $48$ hours \cite{vla3}. In this
experiment we have to take into account several initial routes,
since there are obstacles (islands) each of which doubles the number
of total initial routes \cite{vla2}. The initial routes are plotted
in fig. \ref{fig-init} with the dominant wave direction across the
routes. There are three main regions, the first one has dominant
waves coming from the north (this the region close to Thessaloniki),
the second has dominant waves coming from North-East (the middle
region) and the third has dominant waves coming from South-West
(this the region close to A. Nikolaos).

\begin{figure}
\begin{center}
\includegraphics[width=\textwidth]{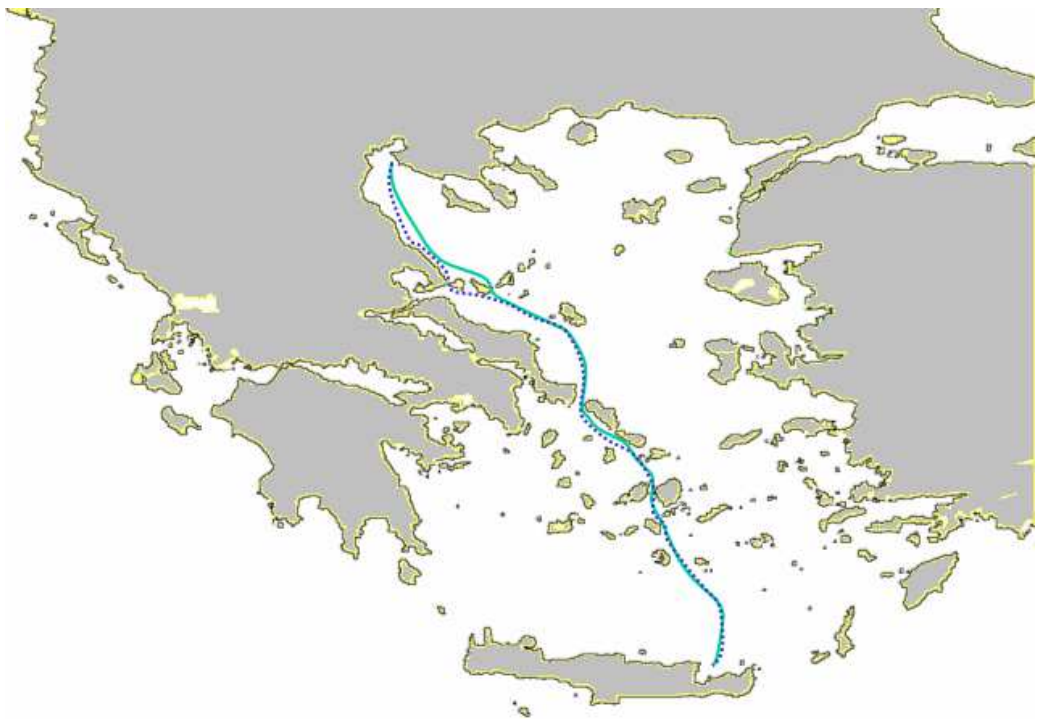}
\end{center}
\caption{Optimal route calculated from the port of Thessaloniki to
A. Nikolaos for the case shown in fig. \ref{fig-init} with the
proposed algorithm (solid line) and with the application of genetic
algorithms (dotted line).}\label{fig-route}
\end{figure}

Fig. \ref{fig-route} shows the calculated optimal route (solid line)
with the one calculated by the application of genetic algorithms
(dotted line). The value of the parameter $\alpha$ was $0.5$. The
two routes agree except for a small part in the beginning of the
voyage which is caused by the property of the route calculated with
the proposed algorithm to be transformed continuously from the
initial route. Nevertheless the difference in the total cost of the
two routes is negligible. An interesting trend of the calculated
route, is that the ship is trying to "\emph{hide}" behind the
obstacles (islands) where the magnitude of the wave height is
decreased dramatically.

\section{Summary and conclusions}
An effective operational algorithm for the calculation of optimal
ship routing has been developed. The algorithm is based on the
simulated annealing technique to search for an optimal solution. The
searching process is accelerated by taking into account a key result,
which enables us to take into account only the variations of the initial route
which are parallel or antiparallel to the direction of the wave
field. Field tests have been performed which exhibit the efficiency
of the proposed algorithm by comparing the calculated routes with
those calculated by exhaustive and time consuming genetic
algorithms.

\section*{Acknowledgments}
This paper is part of the 03ED51 research project, implemented
within the framework of the "\emph{Reinforcement Programme of Human
Research Manpower}" (\textbf{PENED}) and co-financed by National and
Community Funds (25\% from the Greek Ministry of Development-General
Secretariat of Research and Technology and 75\% from E.U.-European
Social Fund).

\end{document}